\documentclass[conference]{IEEEtran}%
\IEEEoverridecommandlockouts

\usepackage{amsfonts}
\usepackage{amsmath}
\usepackage{amssymb}
\usepackage{balance}
\usepackage{ifpdf}
\usepackage{color}
\usepackage[pdftex]{graphicx}

\usepackage{hyperref}

\providecommand{\U}[1]{\protect\rule{.1in}{.1in}}
\makeatletter
\def\ps@headings{\def\@oddhead{\mbox{}\scriptsize\rightmark \hfil \thepage}\def\@evenhead{\scriptsize\thepage \hfil \leftmark\mbox{}}\def\@oddfoot{}\def\@evenfoot{}}
\makeatother
\ifCLASSINFOpdf
\graphicspath{{./pdf/}}
   \DeclareGraphicsExtensions{{.pdf}}
\else
\graphicspath{{./eps/}}
   \DeclareGraphicsExtensions{{.eps}}
\fi
\begin{document}

\title{Guard Zones and the Near-Far Problem in DS-CDMA Ad Hoc Networks \vspace{-0.1cm} }
\author{\IEEEauthorblockN{Don Torrieri\IEEEauthorrefmark{1} and~Matthew C. Valenti\IEEEauthorrefmark{2}}
\IEEEauthorblockA{\IEEEauthorrefmark{1}U.S. Army Research Laboratory, Adelphi, MD, USA.\\
\IEEEauthorrefmark{2}West Virginia University, Morgantown, WV, USA. } 
\thanks{M.C. Valenti's contribution was sponsored by the National Science Foundation under Award No. CNS-0750821 and by the United States Army Research Laboratory under Contract W911NF-10-0109.}
\vspace{-0.6cm} }
\maketitle

\begin{abstract}%

The central issue in direct-sequence code-division multiple-access (DS-CDMA) ad hoc networks is the prevention of a near-far problem. This paper considers two types of guard zones that may be used to control the near-far problem: a fundamental \emph{exclusion zone} and an additional \emph{CSMA guard zone} that may be established by the carrier-sense multiple-access (CSMA) protocol. In the exclusion zone, no mobiles are physically present, modeling the minimum physical separation among mobiles that is always present in actual networks. Potentially interfering mobiles beyond a transmitting mobile's exclusion zone, but within its CSMA guard zone, are deactivated by the protocol. This paper provides an analysis of DS-CSMA networks with either or both types of guard zones. A network of finite extent with a finite number of mobiles is modeled as a uniform clustering process.  The analysis uses a closed-form expression for the outage probability in the presence of Nakagami fading, conditioned on the network geometry.  By using the analysis developed in this paper, the tradeoffs between exclusion zones and CSMA guard zones are explored for DS-CDMA and unspread networks.

\end{abstract}%

\section{Introduction}

Direct-sequence code-division multiple-access (DS-CDMA) ad hoc networks are
realized by using direct-sequence spread-spectrum modulation while the mobiles
or nodes of multiple users simultaneously transmit signals in the same
frequency band. All signals use the entire allocated spectrum, but the
spreading sequences differ. DS-CDMA is advantageous for ad hoc networks
because it eliminates the need for any frequency or time-slot coordination,
imposes no sharp upper bound on the number of mobiles, directly benefits from
inactive terminals in the network, and is capable of exploiting bursty data
traffic, intermittent voice signals, multibeamed arrays, and reassignments to
accommodate variable data rates. Furthermore, DS-CDMA systems are inherently
resistant to interference, interception, and frequency-selective fading.

The central issue in DS-CDMA ad hoc networks is the prevention of a near-far
problem. The solution to the near-far problem in cellular networks is power
control \cite{torr}. However, the absence of a centralized control of an ad
hoc network renders any attempted power control local rather than pervasive.
As an alternative to power control, ad hoc networks typically use \emph{guard
zones} surrounding mobiles to manage interference. This paper considers two
types of guard zones to control the near-far problem: a fundamental
\emph{exclusion zone} and an additional guard zone that may be established by
the carrier-sense multiple-access (CSMA) protocol; i.e., a \emph{CSMA guard
zone}. In the exclusion zone surrounding a mobile, no other mobiles are
physically present. The exclusion zone may be enforced by using information
about the locations of a mobile and its surrounding mobiles and is justified
by the fact that mobiles will always have a minimum physical separation in
actual networks. In the CSMA guard zone \cite{krunz}, \cite{alaw},
\cite{hasan}, which extends beyond the exclusion zone, other mobiles may be
present, but they are deactivated by the protocol. The CSMA guard
zone offers additional near-far protection beyond that offered by the
exclusion zone, but at the cost of reduced network transmission capacity, as
shown subsequently.

In contrast to the existing literature (e.g., \cite{weber}-\cite{bacc2} and
the references therein), a realistic network of finite extent with a finite
number of mobiles and uniform clustering as the spatial distribution is
modeled. The analysis presented in this paper discards previous assumptions
based on stochastic geometry \cite{stoy}, \cite{bacc1} about the spatial
distribution of the mobiles. Both the spatial extent of the network and number
of mobiles are finite, and each mobile has an arbitrary location. The analysis
considers the mobile's duty factor, the shadowing, the possible coordination
among the mobiles (through the use of guard zones), and a Nakagami-m factor
that can vary among the channels from each mobile to the reference receiver. A
recent closed-form expression for the \emph{exact} outage probability \cite{torr3} is
used for a reference receiver conditioned on the network geometry and
shadowing factors. The \emph{spatially averaged outage probability}, defined
as the average over both the mobile locations and the shadowing, is computed
by averaging the conditional outage probability with respect to the spatial
distribution and the distribution of the shadowing. While this expectation is
not generally analytically tractable for finite networks and guard zones, a simple and
efficient Monte Carlo method can be used that involves the random placement of
mobiles and generation of shadowing factors, but does not require the
generation of fading coefficients.


\section{Guard Zones}

\label{Section:GuardZones}

The IEEE 802.11 standard, which is currently the predominant standard for ad
hoc networks, uses CSMA with collision avoidance
in its medium-access control protocol \cite{krunz}, \cite{alaw}. The implementation
entails the exchange of request-to-send (RTS) and clear-to-send (CTS)
handshake packets between a transmitter and receiver during their initial
phase of communication that precedes the subsequent data and acknowledgment
packets. The receipt of the RTS/CTS packets with sufficient power levels by
nearby mobiles causes them to inhibit their own transmissions, which would
produce interference in the receiver of interest. The transmission of separate
CTS packets in addition to the RTS packets decreases the possibility of
subsequent signal collisions at the receiver due to nearby hidden terminals
that do not sense the RTS packets. Thus, the RTS/CTS packets essentially
establish guard zones surrounding a transmitter and receiver, and hence,
prevent a near-far problem except during the initial reception of an RTS
packet. The interference at the receiver is restricted to concurrent
transmissions generated by mobiles outside the guard zones.
The fundamental disadvantage with this CSMA guard zone is that it inhibits other
concurrent transmissions within the zone. 


A more fundamental guard zone, which is
called the \textit{exclusion zone}, is based on the spacing that usually
occurs in actual mobile networks. For instance, when the radios are mounted on
vehicles, there is a need for crash avoidance by maintaining a minimum vehicle
separation. A small exclusion zone maintained by visual sightings exists in
practical networks, but a more reliable and extensive one can be established
by equipping each mobile with a global positioning system (GPS) and
periodically broadcasting each mobile's GPS coordinates. Mobiles that receive
those messages could compare their locations to that in the message and alter
their movements accordingly. The major advantage of the exclusion zone
compared with a CSMA guard zone is that the exclusion zone prevents near-far
problems at receivers while not inhibiting any potential concurrent
transmissions.
Another advantage of an
exclusion zone is enhanced network connectivity because of the inherent
constraint on the clustering of mobiles.

When CSMA is used in a network, the CSMA guard zone will usually encompass the
exclusion zone. Although both zones may cover arbitrary regions, they are
modeled as circular regions in the subsequent examples for computational
convenience, and the region of the CSMA guard zone that lies outside the
exclusion zone is an annular ring. The existence of an annular ring enhances
the near-far protection at the cost of inhibiting potential concurrent
transmissions within the annular ring. The tradeoffs entailed in having the
CSMA guard zone are examined subsequently.

\section{Network Model}

\label{Section:NetworkModel} The network comprises a fixed number of mobiles
in a circular area with radius $r_\mathsf{net}$, although any arbitrary two- or
three-dimensional regions could be considered. A reference receiver is located
inside the circle, a reference transmitter $X_{0}$ is located at distance
$||X_{0}||$ from the reference receiver, and there are $M$ potentially
interfering mobiles $X_{1},...,X_{M}.$ The variable $X_{i}$ represents both
the $i^{th}$ mobile and its location, and $||X_{i}||$ is the distance from
the $i^{th}$ mobile to the receiver. Each mobile uses a single omnidirectional
antenna. The radii of the exclusion zone and the CSMA guard zone are $r_\mathsf{ex}$
and $r_\mathsf{g},$ respectively.

The interfering mobiles are uniformly distributed throughout the network area
outside the exclusion zones, according to a \textit{uniform clustering} model. 
One by one, the location of each $X_i$ is drawn according to a uniform distribution
 within the radius-$r_\mathsf{net}$
circle. However, if an $X_i$ falls within the exclusion zone of a previously
placed mobile, then it has a new random location assigned to  it as many times as necessary
until it falls outside all exclusion zones. Unlike Matern thinning \cite{card},
which silences mobiles without replacing them, uniform clustering maintains a
fixed number of active mobiles.


Since there is no significant advantage to using short direct-sequence
sequences, long spreading sequences are assumed and modeled as random binary
sequences with chip duration $T_{c}$. The processing gain or spreading factor
$G$ directly reduces the interference power.  The multiple-access interference
is assumed to be asynchronous,  and the power from each interfering $X_{i}$ is
further reduced by the chip factor $h(\tau_{i})$, which is a function of the chip waveform and the timing offset
$\tau_{i}$ of $X_{i}$'s spreading sequence relative to that of the desired
signal \cite{torr}.  
In a network of quadriphase direct-sequence systems, a
multiple-access interference signal with power $\mathcal{I}_{i}$ before
despreading is reduced after despreading to the power level $\mathcal{I}_{i} h(\tau_{i})/G,$ 
where \cite{torr}, \cite{torr2}
\begin{equation}
h(\tau_{i})=\frac{1}{T_{c}^{2}}\left[  R_{\psi}^{2}(\tau_{i})+R_{\psi}%
^{2}(T_{c}-\tau_{i})\right]
\end{equation}
and $R_{\psi}(\tau_{i})$ is the partial autocorrelation for the normalized
chip waveform. Thus, the interference power is effectively reduced by the
factor $G_i= G/h(\tau_{i}).$ 
Assuming a rectangular chip waveform and that $\tau_{i}$ has a uniform distribution over [0, $T_{c}],$  
the expected value of $h(\tau_{i})$ is 2/3.

After despreading, the power of $X_{i}$'s signal at the reference receiver is
\begin{equation}
\rho_{i}= \tilde{P}_{i}g_{i}10^{\xi_{i}/10}f\left(  ||X_{i}|| \right)  \label{power}%
\end{equation}
where $\tilde{P}_{i}$ is the received power 
at a reference distance $d_0$ (assumed to be sufficiently far that the signals are in the far
field) after despreading when fading and shadowing are absent,
$g_{i}$ is the power gain due to fading, $\xi_{i}$ is a shadowing coefficient, 
and $f(\cdot)$ is a path-loss function. 
The path-loss function is expressed as the power law
\begin{equation}
f\left(  d\right)  =\left(  \frac{d}{d_0}\right)  ^{-\alpha}\hspace
{-0.45cm}, \, \, \text{ \ }d\geq d_0 \label{pathloss}%
\end{equation}
where $\alpha\geq2$ is the path-loss exponent. It is assumed that $r_\mathsf{ex} \geq d_0.$   The \{$g_{i}\}$ are
independent with unit-mean, but are not necessarily identically distributed;
i.e., the channels from the different $\{X_{i}\}$ to the reference receiver
may undergo fading with different distributions. For analytical tractability
and close agreement with measured fading statistics, Nakagami fading is
assumed, and $g_{i}=a_{i}^{2}$, where $a_{i}$ is Nakagami with parameter
$m_{i}$. When the channel between $X_{i}$ and the reference receiver undergoes
Rayleigh fading, $m_{i}=1$ and the corresponding $g_{i}$ is exponentially
distributed. In the presence of log-normal shadowing, the $\{\xi_{i}\}$ are
independent zero-mean Gaussian with variance $\sigma_{s}^{2}$. For ease of
exposition, it is assumed that the shadowing variance is the same for the
entire network, but the results may be easily generalized to allow for different
shadowing variances over parts of the network. In the absence of shadowing, $\xi_{i}=0$.

It is assumed that the \{$g_{i}\}$ remain fixed for the duration of a time
interval, but vary independently from interval to interval (block fading).
With probability $p_{i}$, the $i^{th}$ mobile transmits in the same time
interval as the desired signal. The $\{p_{i}\}$ can be used to model
voice-activity factors, controlled silence, or failed link transmissions and
the resulting retransmission attempts. The $\{p_{i}\}$ need not be the same;
for instance, CSMA protocols can be modeled by setting $p_{i}=0$ when a mobile
lies within the CSMA guard zone of another active mobile, which is equivalent
to implementing Matern thinning within the annular ring corresponding to that
guard zone.

The instantaneous signal-to-interference-and-noise ratio (SINR) at the
receiver is given by: 
\begin{equation}
\gamma=\frac{\rho_{0}}{\displaystyle{\mathcal{N}}+\sum_{i=1}^{M}I_{i}\rho_{i}}
\label{SINR1}%
\end{equation}
where $\rho_{0}$ is the received power of the desired signal, $\mathcal{N}$ is the
noise power, and the indicator $I_{i}$ is a Bernoulli random variable with
probability $P[I_{i}=1]=p_{i}$ and $P[I_{i}=0]=1-p_{i}$.

Since the despreading does not affect the desired-signal power, the substitution of  (\ref{power}) and (\ref{pathloss}) into (\ref{SINR1}) yields
\begin{equation}
\gamma=\frac{g_{0}\Omega_{0}}{\displaystyle\Gamma^{-1}+\sum_{i=1}^{M}%
I_{i}g_{i}\Omega_{i}} \label{SINR2}%
\end{equation}
where 
\begin{equation}
\Omega_{i}=%
\begin{cases}
10^{\xi_{0}/10}||X_{0}||^{-\alpha} & i=0\\
\displaystyle\frac{{P}_{i}}{G_i P_{0}}10^{\xi_{i}/10}||X_{i}||^{-\alpha} & i>0
\end{cases}
\label{eqn:omega}%
\end{equation}
is the normalized power of $X_{i}$, ${P}_{i}$ is the received power
at the reference distance $d_0$ before despreading when fading and shadowing
are absent, and $\Gamma=d_0^{\alpha}P_{0}/\mathcal{N}$ is the SNR when the reference
transmitter is at unit distance from the reference receiver and fading and
shadowing are absent.

\section{Outage Probability}

\label{Section:Outage} Let $\beta$ denote the minimum SINR required for
reliable reception and $\boldsymbol{\Omega}=\{\Omega_{0},...,\Omega_{M}\}$
represent the set of normalized powers. An \emph{outage} occurs when the SINR
falls below $\beta$. Conditioning on $\boldsymbol{\Omega}$, the outage
probability is
\begin{equation}
\epsilon=P\left[  \gamma\leq\beta\big|\boldsymbol{\Omega}\right]  .
\label{Equation:Outage1}%
\end{equation}
The outage probability depends on the network geometry and shadowing factors,
which have dynamics over much slower timescales than the fading. In contrast,
the outage probability of \cite{weber} is not conditioned on
$\boldsymbol{\Omega}$ and therefore is unable to quantify the outage
probability of any specific network geometry. By defining a variable
\vspace{-0.25cm}
\begin{equation}
\mathsf{Z}=\beta^{-1}g_{0}\Omega_{0}-\sum_{i=1}^{M}I_{i}g_{i}\Omega_{i}
\label{eqn:z}%
\end{equation}
\vspace{-0.05cm}
the conditional outage probability may be expressed as
\begin{equation}
\epsilon=P\left[  \mathsf{Z}\leq\Gamma^{-1}\big|\boldsymbol{\Omega}\right]
=F_{\mathsf{Z}}\left(  \Gamma^{-1}\big|\boldsymbol{\Omega}\right)
\label{Equation:OutageCDF}%
\end{equation}
which is the cumulative distribution function (cdf) of $\mathsf{Z}$
conditioned on $\boldsymbol{\Omega}$ and evaluated at $\Gamma^{-1}$.

Define $\bar{F}_{\mathsf{Z}}(z)=1-F_{\mathsf{Z}}(z)$ to be the complementary
cdf (ccdf) of $\mathsf{Z}$. Restricting the Nakagami parameter $m_{0}$ of the channel
between the reference transmitter and  receiver to be integer-valued,
the ccdf of $\mathsf{Z}$ conditioned on $\boldsymbol{\Omega}$ is
\cite{torr3}
\begin{equation}
\bar{F}_{\mathsf{Z}}\left(  z\big|\boldsymbol{\Omega}\right)  =e^{-\beta_{0}%
z}\sum_{s=0}^{m_{0}-1}{\left(  \beta_{0}z\right)  }^{s}\sum_{t=0}^{s}%
\frac{z^{-t}H_{t}(\boldsymbol{\Psi})}{(s-t)!} \label{NakagamiCond}%
\end{equation}
where $\beta_{0}=\beta m_{0}/\Omega_{0}$,
\begin{align}
\Psi_{i}  &  =\left(  \beta_{0}\frac{\Omega_{i}}{m_{i}}+1\right)  ^{-1}%
\hspace{-0.5cm},\hspace{1cm}\mbox{for $i=\{1,...,M\}$, }\label{Psi}\\
H_{t}(\boldsymbol{\Psi})  &  =\mathop{ \sum_{\ell_i \geq 0}}_{\sum_{i=0}%
^{M}\ell_{i}=t}\prod_{i=1}^{M}{\mathsf G}_{\ell_{i}}(\Psi_{i}), \label{Hfunc}%
\end{align}
the summation in (\ref{Hfunc}) is over all sets of indices that sum to $t$,
and
\vspace{-0.35cm}
\begin{equation}
\mathsf{G}_{\ell}(\Psi_{i})=%
\begin{cases}
1-p_{i}(1-\Psi_{i}^{m_{i}}) & \mbox{for $\ell=0$}\\
\frac{p_{i}\Gamma(\ell+m_{i})}{\ell!\Gamma(m_{i})}\left(  \frac{\Omega_{i}%
}{m_{i}}\right)  ^{\ell}\Psi_{i}^{m_{i}+\ell} & \mbox{for $\ell>0$.}
\end{cases}
\label{Gfunc}%
\end{equation}

\section{Examples}

\begin{figure}[t]
\centering
\vspace{-0.1cm} \includegraphics[width=6.5cm]{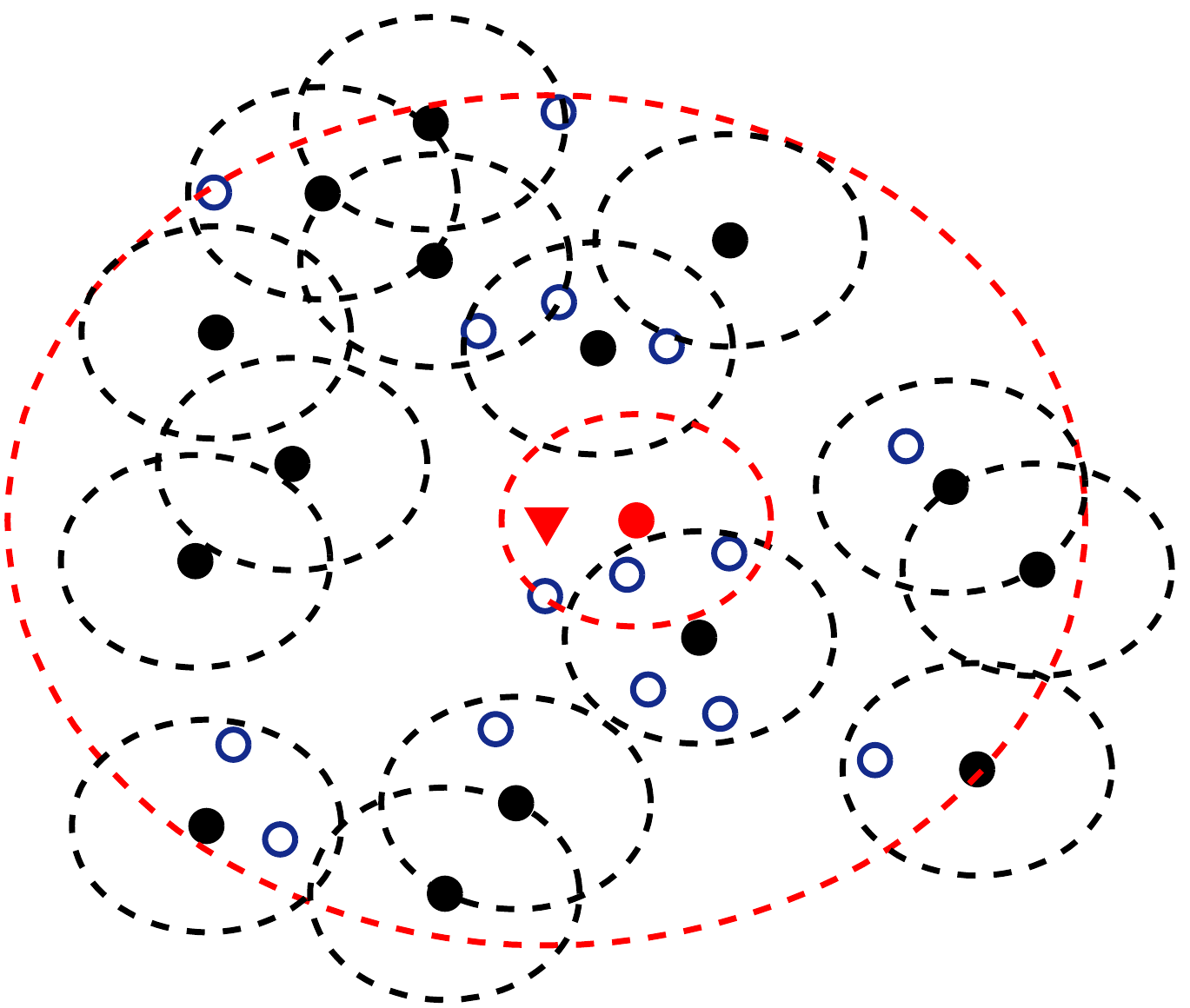} \vspace{-0.25cm} \caption{
Example network realization. The reference receiver (indicated by the {\color{red}
$\blacktriangledown$}) is placed at the origin, and the reference
transmitter is at $X_{0}=1/6$ (indicated by the {\color{red} $\bullet$} to its right). 
$M=30$ mobiles are placed according to the uniform clustering model, each 
with an exclusion zone (not shown) of radius $r_\mathsf{ex}=1/12$. Active mobiles are indicated by filled
circles while deactivated mobiles are indicated by unfilled circles. A guard
zone of radius $r_\mathsf{g}=1/4$ surrounds each active mobile, as depicted by dashed
circles. When CSMA guard zones are used, the other mobiles within the guard
zone of an active mobile are deactivated. \vspace{-0.65cm} }%
\label{Figure:Fig1}%
\end{figure}

\label{Section:Examples} In the examples, all distances are normalized to the
network radius so that $r_\mathsf{net} = 1$, and the reference receiver is located at the center of 
the network unless otherwise stated.  As shown in Fig. \ref{Figure:Fig1}, the reference transmitter is at coordinate $X_0 = 1/6$ and $M=30$ potentially interfering mobiles are placed
according to a uniform clustering model with a radius $r_\mathsf{ex}=1/12$ exclusion zone surrounding each mobile.
Once the mobile locations $\{X_{i}\}$ are realized, the $\{\Omega_{i}\}$ are determined by assuming a path-loss
coefficient $\alpha=3.5$ and a common transmit power (${P}_{i}/P_{0}=1$ for
all $i$).  Both unshadowed and shadowed environments are considered. 
Although the model permits nonidentical $G_i$, it is assumed that $G_i$ is a constant equal to $G_\mathsf{e}$ (the {\em effective} spreading gain) for all interference signals. Both spread and unspread  systems are considered, with $G_\mathsf{e}=1$ for the unspread system and $G_\mathsf{e}=48$
for the spread system, corresponding to a typical direct-sequence waveform
with $G=32$ and $h(\tau_{i})=2/3$. The value of $p_{i}=0.5$ for all active $X_i$, and the Nakagami parameters
are $m_{0}=3$ for the desired signal and $m_{i}=1$ for the interfering mobiles
(i.e., Rayleigh fading). Using two different Nakagami factors (i.e.,
\emph{mixed} fading) is justified by the fact that the reference transmitter
is usually within the line-of-sight of the receiver while the interferers are
not and, therefore, are subject to more severe fading. The SINR threshold is
$\beta=0$ dB, which corresponds to the unconstrained AWGN capacity limit for a
rate-$1$ channel code, assuming complex-valued inputs.

\begin{figure}[t]
\centering
\vspace{-0.1cm}
\includegraphics[width=8.75cm]{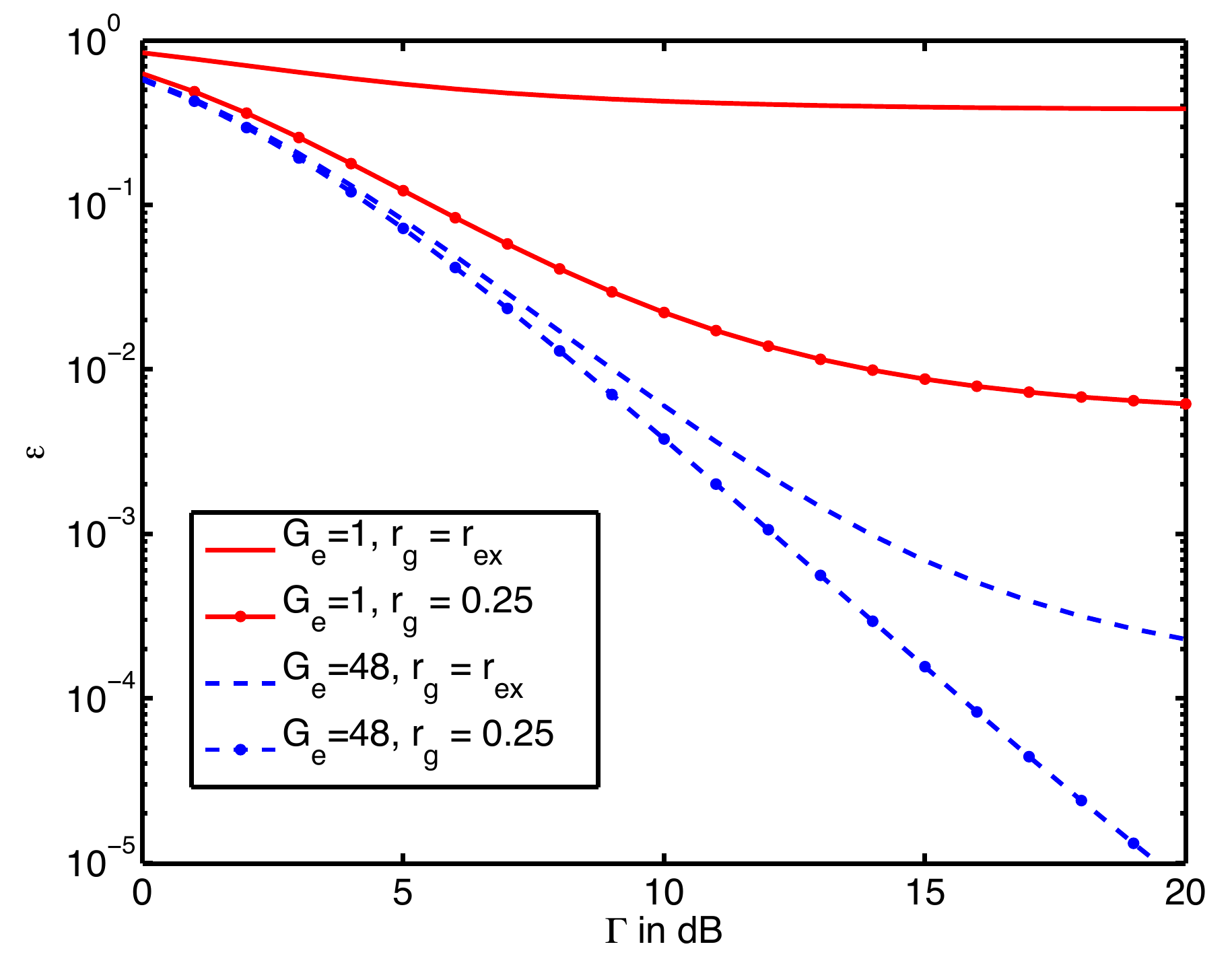} 
\vspace{-0.8cm} \caption{
Outage probability as a function of SNR, conditioned on the network
shown in Fig. 1 with mixed fading and no shadowing. Dashed lines are with spreading ($G_\mathsf{e}=48$) while solid
lines are without. Curves marked with $\bullet$ indicate performance with a
CSMA guard zone of radius $r_\mathsf{g}=1/4$ while curves without markers
indicate performance without CSMA guard zones. 
\vspace{-0.65cm} }%
\label{Figure:Fig2}%
\end{figure}

\textbf{Example \#1.} Suppose that CSMA is not used, and therefore there is no
guard zone beyond the fundamental exclusion zone; i.e., $r_\mathsf{g}=r_\mathsf{ex}$. All 30
potentially interfering mobiles in Fig. \ref{Figure:Fig1} remain active and 
contribute to the overall interference at the
reference receiver. The outage probability for this network is shown in Fig.
\ref{Figure:Fig2} by the two curves without markers, corresponding to the
unspread ($G_\mathsf{e}=1$) and direct-sequence spread ($G_\mathsf{e}=48$) networks. Without
direct-sequence spreading, the outage probability is quite high, for instance
$\epsilon=0.4$ at $\Gamma=20$ dB. Spreading reduces the outage probability by
about three orders of magnitude at high SNR, although this comes at the cost
of increased required bandwidth.

\textbf{Example \#2.} The outage probability can be reduced by using CSMA to
impose guard zones beyond the fundamental exclusion zone. Suppose that a guard
zone of radius $r_\mathsf{g}=1/4$ is used, as shown in Fig. \ref{Figure:Fig1}.
Potentially interfering mobiles are deactivated according the following
procedure, which is equivalent to Matern thinning. First, the reference
transmitter $X_{0}$ is activated. Next, each potentially interfering mobile is
considered in the order it was placed. For each mobile, a check is made to see
if it is in the guard zone of a previously placed active mobile. Since mobiles
are indexed according to the order of placement, $X_{1}$ is first considered
for possible deactivation; if it lies in the guard zone of $X_{0}$, it is
deactivated, and otherwise it is activated. The process repeats for each
subsequent $X_{i}$, deactivating it if it falls within the guard zone of any
active $X_{j}$, $j<i$, or otherwise activating it.


In Fig. \ref{Figure:Fig1}, active mobiles are indicated by filled circles and deactivated mobiles are
indicated by unfilled circles. The guard zone around each active mobile is
indicated by a dashed circle of radius $r_\mathsf{g}$. The reference receiver has not
been assigned a CSMA guard zone, which reflects the fact that it has none
while it is receiving the initial RTS. In the given example, 15 mobiles have
been deactivated, while the remaining 15 mobiles remain active. The outage
probability of the network with deactivated mobiles is shown in Fig.
\ref{Figure:Fig2} for $G_\mathsf{e}=\{1,48\}$ by the two curves with markers. The
performance of the unspread network improves dramatically with a guard zone,
being reduced by two orders of magnitude at high SNR. The DS-CDMA network,
which already had superior performance without CSMA, has improved performance,
but the improvement only becomes significant at a high SNR. 

\section{Spatial Averaging}

\label{Section:SpatialAveraging}

Because it is conditioned on ${\boldsymbol{\Omega}}$, the outage probability
will vary from one network realization to the next. The conditioning on
${\boldsymbol{\Omega}}$ can be removed by averaging $\bar{F}_{\mathsf{Z}%
}(z|\boldsymbol{\Omega})$ 
over the network geometry. This can be done analytically
only under certain limitations \cite{torr3}. However, the outage probability
can be estimated through Monte Carlo simulation by generating many different
${\boldsymbol{\Omega}}$, computing the outage probability of each, and taking
the numerical average.   Note that generating each ${\boldsymbol{\Omega}}$ involves not
only placing the mobiles according to the uniform clustering modeling, but 
also realizing the shadowing and deactivating mobiles that lie within the CSMA 
guard zone of active mobiles.

\begin{figure}[t]
\centering
\vspace{-0.1cm} \includegraphics[width=8.75cm]{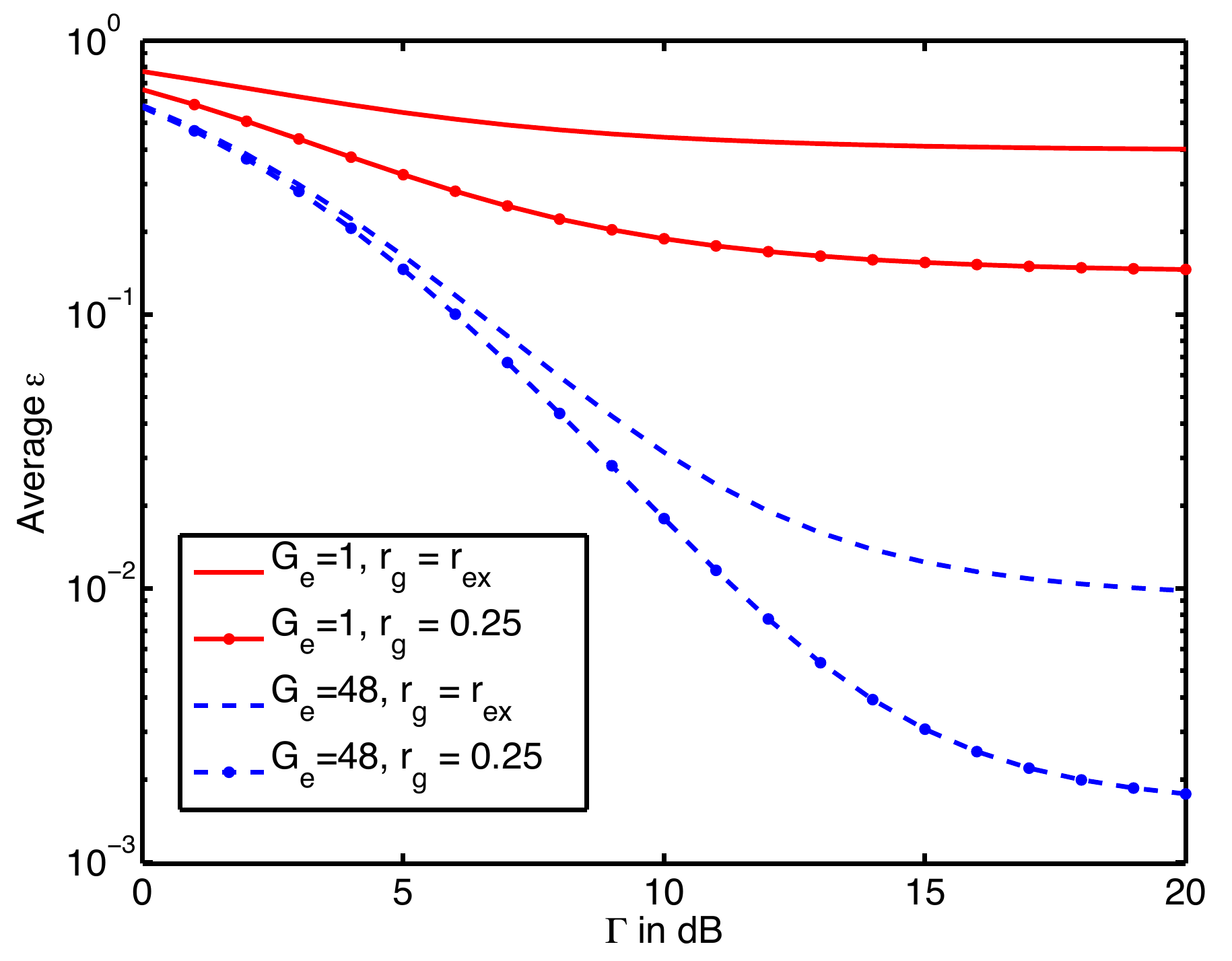} 
\vspace{-0.8cm}
\caption{
Outage probability in mixed fading as a function of SNR averaged over
$10,000$ networks drawn under the same conditions as the one shown in Fig.
1. For each network, shadowing with $\sigma_{s}=8$ dB is applied. \vspace{-0.6cm}}%
\label{Figure:Fig3}%
\end{figure}

As an example, Fig. \ref{Figure:Fig3} shows the average outage probability computed over a set of
$10,000$ networks, each generated the same way as the Examples given in Section \ref{Section:Examples}. 
For $r_\mathsf{g}=r_\mathsf{ex}$, each network was generated
by placing $M=30$ mobiles with exclusion zone $r_\mathsf{ex}=1/12$ according to a uniform clustering model. 
For $r_\mathsf{g} = 1/4$, the same set of networks was used that was already generated with $r_\mathsf{ex}=1/12$, 
but in each network, potentially interfering mobiles were deactivated if they were in the guard zone
of an active mobile. Log-normal shadowing with $\sigma_{s} = 8$ dB was
applied, and all other system parameters are the same used in Section \ref{Section:Examples}.

In a finite network, the average outage probability depends on the location of
the reference receiver. Rather than leaving the reference receiver at the
origin, Table 1 explores the change in performance when the reference receiver
moves from the center of the radius-$r_\mathsf{net}$ circular network to its
perimeter. The SNR is $\Gamma=10$ dB, all channels undergo
mixed fading ($m_{0}=3$ and $m_{i}=1,i\geq1$) with lognormal shadowing
($\sigma_{s}=8$ dB), $M=30$, $\beta=0$ dB, and $p_{i}=0.5$ for the
active interferers. For each set of values of the parameters $G_\mathsf{e}$, $\alpha,$
$r_\mathsf{ex},$ and $r_\mathsf{g},$ the average outage probability at the network center
$\bar{\epsilon}_{c}$ and at the network perimeter $\bar{\epsilon}_{p}$ were
computed by averaging over $10,000$ network realizations. The
interferers were placed according to the uniform clustering model, and the
reference transmitter was placed at distance $||X_{0}||=1/6$ from the
reference receiver. Two values of each parameter were considered:
$G_\mathsf{e}=\{1,48\},$ $\alpha=\{3,4\},$ $r_\mathsf{ex}=\{0,1/12\},$ $r_\mathsf{g}=\{1/12,1/4\}.$
The table indicates that $\bar{\epsilon}_{p}$ is considerably less than
$\bar{\epsilon}_{c}$ in the finite network. This result cannot be predicted by
the traditional infinite-network model, which cannot differentiate between
$\bar{\epsilon}_{c}$ and $\bar{\epsilon}_{p}$. The reduction in outage
probability is more significant for the unspread network and is less
pronounced with increasing $G_\mathsf{e}$. Given a small exclusion zone with
$r_\mathsf{ex}=1/12$, both direct-sequence spreading and CSMA make the average
performance less dependent on the location of the reference receiver inside
the network.

\begin{table}[ptb]
\caption{Average outage probability when the receiver is at the center
($\bar{\epsilon}_{c}$) and at the perimeter ($\bar{\epsilon}_{p}$) of the
network.}%
\vspace{-0.3cm}
\centering
\par%
\begin{tabular}
[c]{|c|c|c|c|c|c|}\hline
$G_\mathsf{e}$ & $\alpha$ & $r_\mathsf{ex}$ & $r_\mathsf{g}$ & $\bar{\epsilon}_{c}$ &
$\bar{\epsilon}_{p}$\\\hline
1 & 3 & 0 & 1/12 & 0.5298 & 0.3056\\\cline{4-6}
&  &  & 1/4 & 0.2324 & 0.1683\\\cline{3-6}
&  & 1/12 & 1/12 & 0.5234 & 0.2592\\\cline{4-6}
&  &  & 1/4 & 0.2256 & 0.1528\\\cline{2-6}
& 4 & 0 & 1/12 & 0.4129 & 0.2388\\\cline{4-6}
&  &  & 1/4 & 0.1453 & 0.1228\\\cline{3-6}
&  & 1/12 & 1/12 & 0.3869 & 0.1774\\\cline{4-6}
&  &  & 1/4 & 0.1313 & 0.1026\\\cline{1-6}%
48 & 3 & 0 & 1/12 & 0.0644 & 0.0391\\\cline{4-6}
&  &  & 1/4 & 0.0181 & 0.0172\\\cline{3-6}
&  & 1/12 & 1/12 & 0.0308 & 0.0199\\\cline{4-6}
&  &  & 1/4 & 0.0173 & 0.0165\\\cline{2-6}
& 4 & 0 & 1/12 & 0.0842 & 0.0494\\\cline{4-6}
&  &  & 1/4 & 0.0177 & 0.0174\\\cline{3-6}
&  & 1/12 & 1/12 & 0.0335 & 0.0209\\\cline{4-6}
&  &  & 1/4 & 0.0165 & 0.0163\\\cline{1-6}%
\end{tabular}
\vspace{-0.65cm}\end{table}

\section{Transmission Capacity}

\label{Section:TC} While the outage probability is improved with CSMA due to
the deactivation of potential interferers, the overall network becomes less
efficient due to the suppression of transmissions. The network efficiency can
be quantified by the \emph{transmission capacity} (TC) \cite{weber}: 
\begin{equation}
\tau=(1-\epsilon)\lambda b
\end{equation}
where $\lambda$ is the network density, which is
the number of active mobiles per unit area, and $b$ is the link throughput in the absence of an outage, in units of bps.
The TC represents the network throughput per unit area.  Increasing the size of the guard zone generally reduces TC due to fewer simultaneous transmissions.

For a given value of $M$, the density without a CSMA guard zone remains fixed
since all interferers remain active. However, with a CSMA guard zone, the
number of active interferers $m$ is random with a value that depends on the
value of $r_\mathsf{g}$, the locations of the interferers, and their order of
placement, which affects how they are deactivated. As with outage probability,
Monte Carlo simulation can be used to estimate TC. 

\subsection{Effect of Transmitter Distance}

\begin{figure}[t]
\centering
\includegraphics[width=8.5cm]{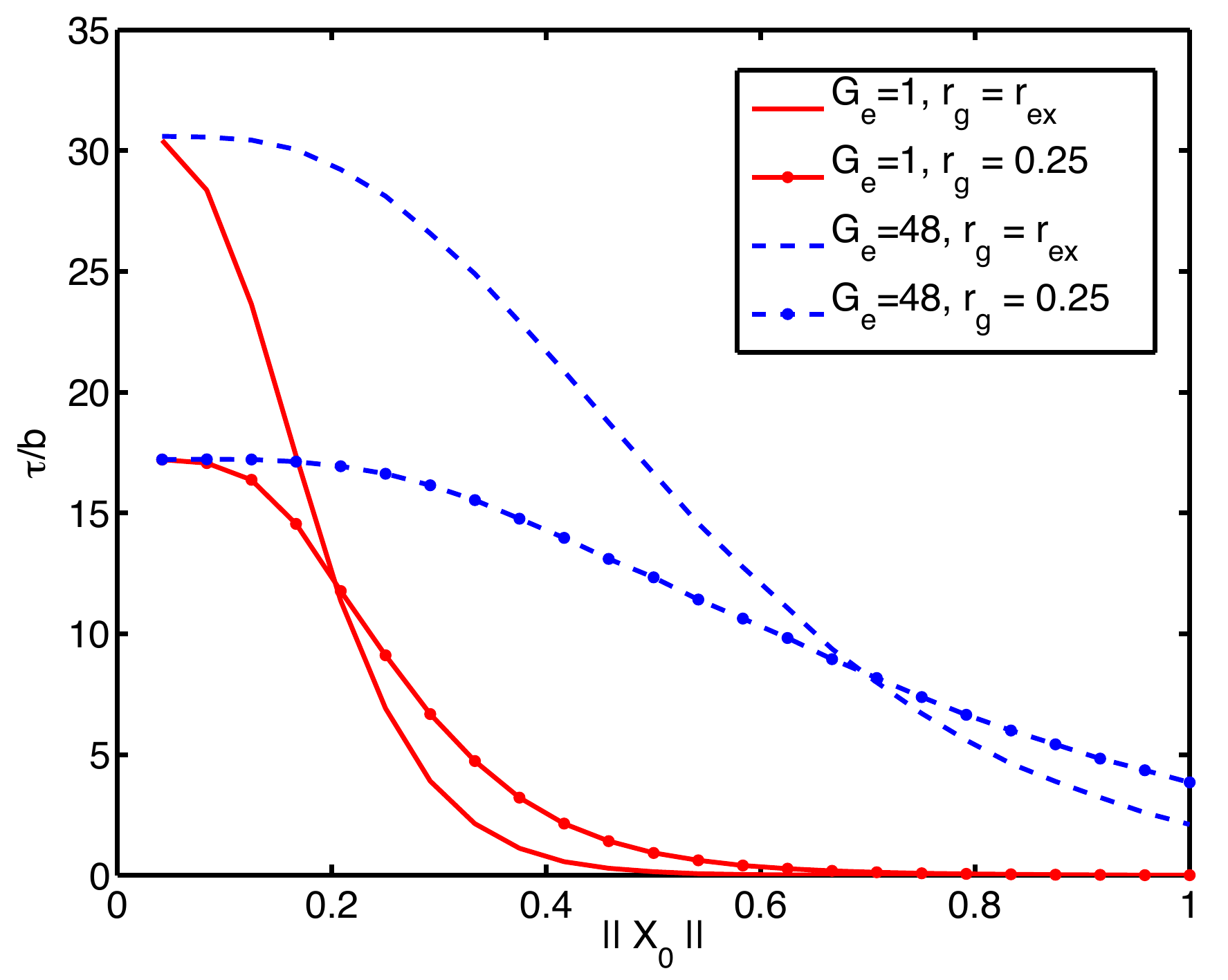} 
\vspace{-0.45cm} 
\caption{Transmission capacity for different transmit distances $||X_0||$. 
The channel model is the same as in Fig. \ref{Figure:Fig3} at $\Gamma=10$ dB. }%
\vspace{-0.5cm}
\label{Figure:MoveTX}%
\end{figure}

The previous examples assume that the distance between the reference
transmitter and receiver is fixed. However, performance will depend on this
distance. Fig. \ref{Figure:MoveTX} shows $\tau/b,$ the normalized TC as a function of the distance 
$||X_0||$ between the reference transmitter and receiver. Except for the value of $||X_0||$, 
the plot was created using the same conditions and parameters used to produce
Fig. \ref{Figure:Fig3} at SNR $\Gamma=10$ dB.  It is
observed that increasing the transmission distance reduces the TC due to the
increase in the number of interferers that are closer to the receiver than the
reference transmitter. However, the degradation in performance is more gradual
with the spread system than the unspread one. The degradation is made more
gradual by the use of CSMA, and at extreme distances, the spread system with
CSMA outperforms the spread system that does not use CSMA.

\subsection{Effect of $r_\mathsf{ex}$ and $r_\mathsf{g}$}
\begin{figure}[t]
\centering
\vspace{-0.1cm}
\includegraphics[width=8.75cm]{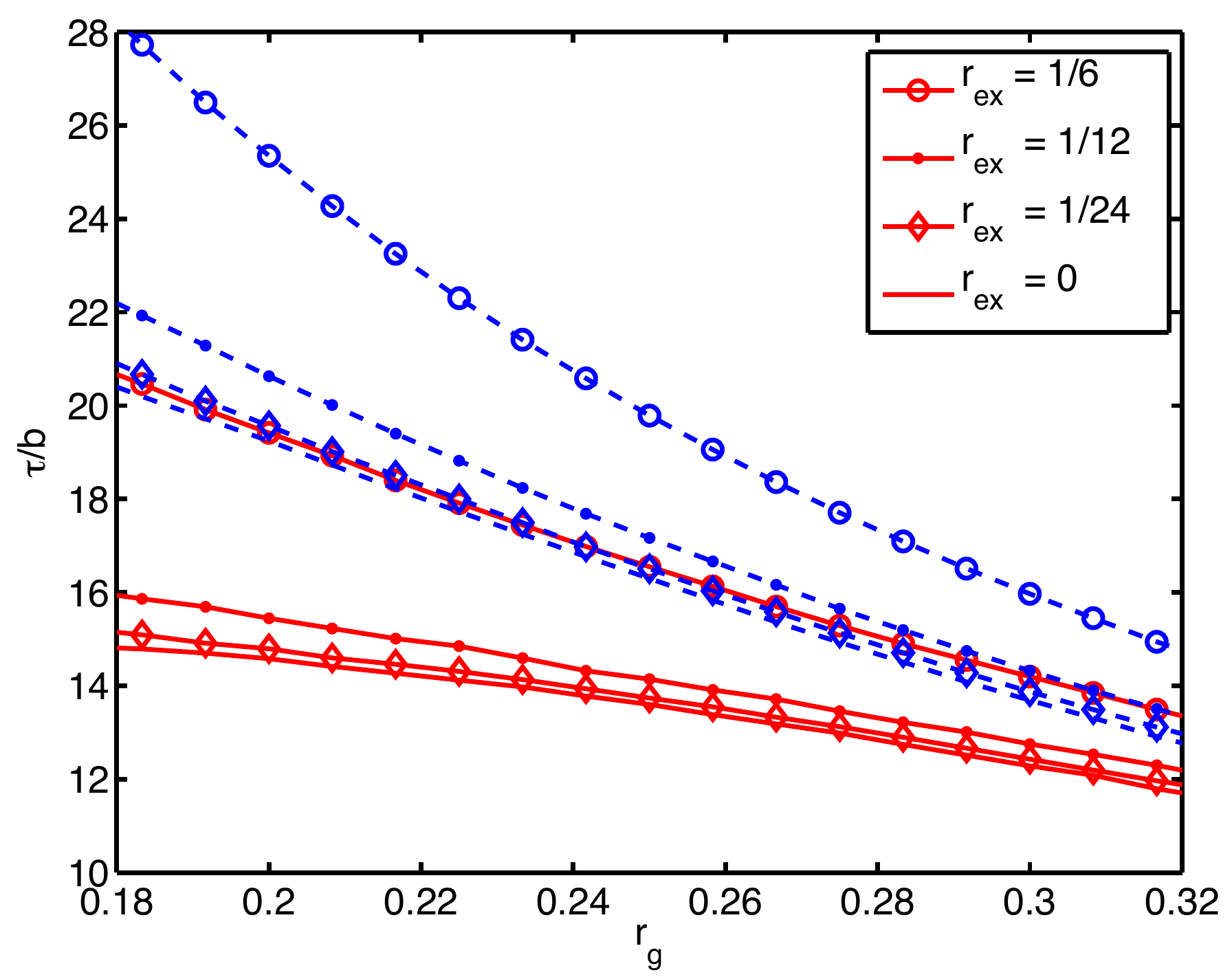} 
\vspace{-0.75cm}%
\caption{Transmission capacity as a function of $r_\mathsf{g}$ for several values of
$r_\mathsf{ex}$. Dashed lines are with spreading ($G_\mathsf{e}=48$) while solid lines are
without ($G_\mathsf{e}=1$). The channel model is the same as in Fig.
\ref{Figure:Fig3} at $\Gamma=10$ dB. }%
\vspace{-0.6cm}
\label{Figure:Fig6}%
\end{figure}

The exclusion-zone radius $r_\mathsf{ex}$ and guard-zone radius $r_\mathsf{g}$ influence the
TC of the network. To study this relationship, the TC was determined over a
range of $r_\mathsf{ex}$ and $r_\mathsf{g}$. With the exception of the values of $r_\mathsf{ex}$
and $r_\mathsf{g}$, the same conditions and parameters used to produce Fig.
\ref{Figure:Fig3} are again used, with $\Gamma=10$ dB. In
Fig. \ref{Figure:Fig6}, the guard-zone radius $r_\mathsf{g}$ was varied over $1/6\leq
r_\mathsf{g}\leq1/3$. For each $r_\mathsf{g}$, the transmission capacity was found for no
exclusion zone ($r_\mathsf{ex}=0$) and exclusion zones with $r_\mathsf{ex}%
=\{1/24,1/12,1/6\}$. While transmission capacity was better with larger
$r_\mathsf{ex}$, the transmission capacity of larger values of $r_\mathsf{ex}$ diminishes
quickly with increasing $r_\mathsf{g}$. Although the reference transmitter was placed
at distance $||X_0||=1/6$ from the reference receiver in Fig.
\ref{Figure:Fig6}, the results remain qualitatively the same for a wide range
of values of $||X_0||.$ 

\subsection{Effect of the Number of Interferers}
\begin{figure}[t]
\centering
\vspace{-0.1cm} 
\includegraphics[width=8.75cm]{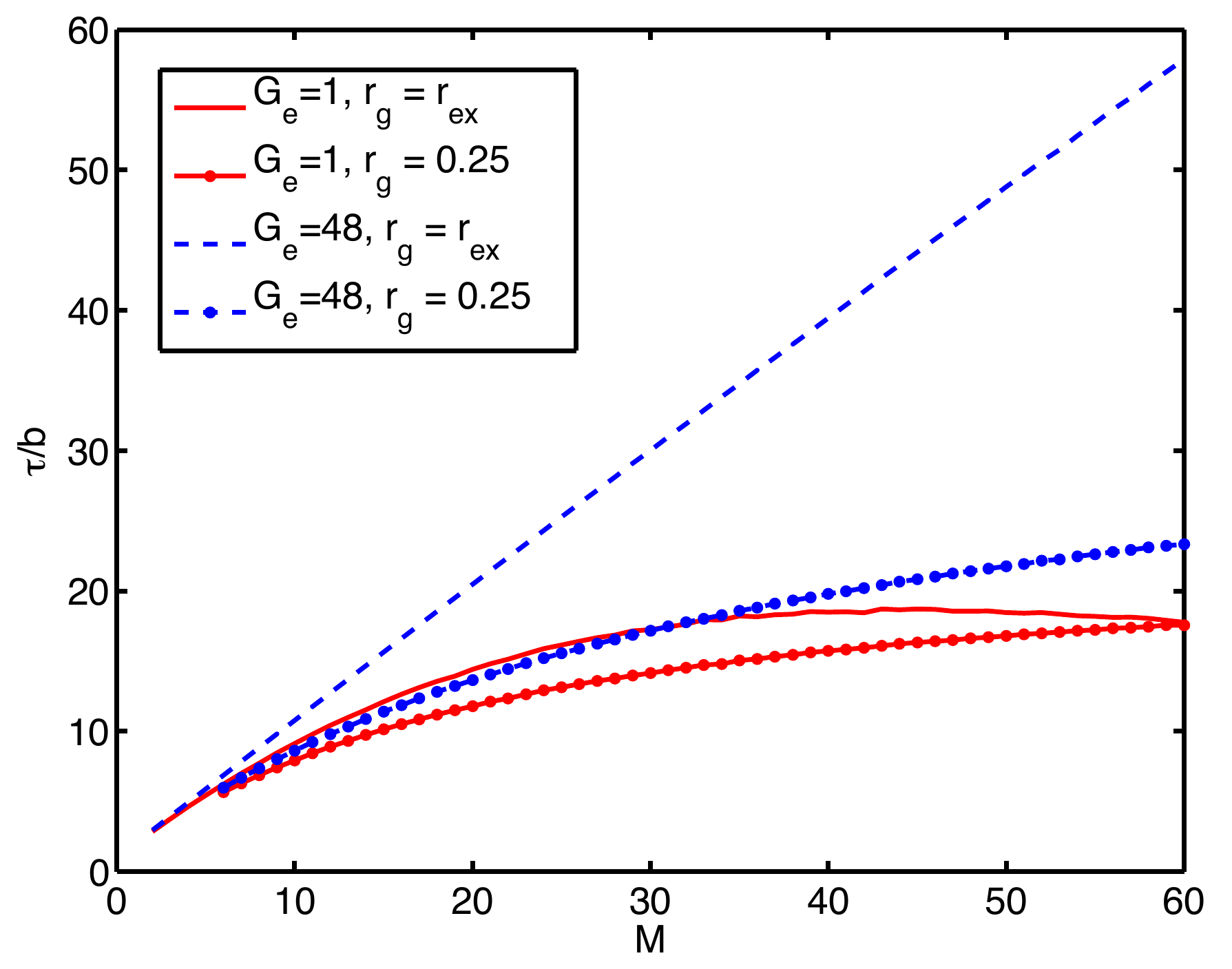} 
\vspace{-0.75cm}%
\caption{Transmission capacity as a function of the number of placed
mobiles $M$. An exclusion zone of $r_\mathsf{ex} = 1/12$ is used.
Performance is shown both with spreading ($G_\mathsf{e}=48$) and without spreading
($G_\mathsf{e}=1$), and both with a CSMA guard zone ($r_\mathsf{g}=1/4$) and without an
additional CSMA guard zone ($r_\mathsf{g}=r_\mathsf{ex}$). The channel model is the same as
in Fig. \ref{Figure:Fig3} at $\Gamma= 10$ dB. }%
\label{Figure:Fig7}%
\vspace{-0.65cm}
\end{figure}

In the previous figures, a fixed number of interferers ($M=30$) was placed
prior to CSMA deactivation. In Fig. \ref{Figure:Fig7}, the number of
interferers is varied from 2 to 60, and the transmission capacity computed for
each value of $M$ for a representative set of $G_\mathsf{e}$ and $r_\mathsf{g}$. Except for
the value of $M$, the plot was created using the same conditions and
parameters used to produce Fig. \ref{Figure:Fig3} at
$\Gamma=10$ dB. As observed previously for just $M=30$, the transmission
capacity of the DS-CDMA network is higher than that of the unspread network, and using
a CSMA guard zone reduces transmission capacity. The transmission capacity of
the DS-CDMA network increases roughly linearly with $M$ in the absence of a
guard zone while the increase is sublinear for the unspread network or when a
guard zone is used. At $M=60$, the transmission capacity of the unspread
network is approximately the same with and without a guard zone.

\section{Conclusions}

\label{Section:Conclusions}

The analysis developed in this paper allows the tradeoffs between exclusion
zones and CSMA guard zones to be explored for DS-CDMA and unspread networks.
The advantage of an
exclusion zone over a CSMA guard zone is that since the network is not
thinned, the number of active mobiles remains constant and higher transmission
capacities can be achieved. If the processing gain is sufficiently large, a
CSMA guard zone is largely ineffective in that it only slightly improves the
outage probability at the cost of a considerable decrease in the transmission capacity.

\end{document}